\documentclass[runningheads]{llncs}
\usepackage[T1]{fontenc}
\usepackage{graphicx}
\usepackage{hyperref}
\usepackage{color}

\usepackage{amsmath}
\usepackage{booktabs}
\begin{document}
\title{Stat-weight: Improving the Estimator of Interleaved Methods Outcomes with Statistical Hypothesis Testing}
\titlerunning{Improving Interleaved Methods with Statistical Hypothesis Testing}
%
\author{Alessandro Benedetti\orcidID{0000-0002-8781-8619} \and
\\Anna Ruggero\orcidID{0000-0003-1733-1132}}
%
%
\institute{Sease Ltd., London, UK\\
\email{a.benedetti@sease.io},
\email{a.ruggero@sease.io}\\
}
\maketitle              
\begin{abstract}
Interleaving is an online evaluation approach for information retrieval systems that compares the effectiveness of ranking functions in interpreting the users' implicit feedback.
Previous work such as Hofmann et al (2011) \cite{hofmann2011probabilistic} has evaluated the most promising interleaved methods at the time, on uniform distributions of queries.
In the real world, ordinarily, there is an unbalanced distribution of repeated queries that follows a long-tailed users' search demand curve.
The more a query is executed, by different users (or in different sessions), the higher the probability of collecting implicit feedback (interactions/clicks) on the related search results.
This paper first aims to replicate the Team Draft Interleaving accuracy evaluation on uniform query distributions and then focuses on assessing how this method generalizes to long-tailed real-world scenarios.
The reproducibility work raised interesting considerations on how the winning ranking function for each query should impact the overall winner for the entire evaluation.
Based on what was observed, we propose that not all the queries should contribute to the final decision in equal proportion.
As a result of these insights, we designed two variations of the $\Delta_{AB}$ score winner estimator that assign to each query a credit based on statistical hypothesis testing.
To replicate, reproduce and extend the original work, we have developed from scratch a system that simulates a search engine and users' interactions from datasets from the industry.
Our experiments confirm our intuition and show that our methods are promising in terms of accuracy, sensitivity, and robustness to noise.

\end{abstract}
\section{Introduction}
Online evaluation is used to estimate the best ranking function for an information retrieval system, directly addressing a live instance with real users and data.
It compares ranking functions through the interpretation of the users' behavior, represented by the collected interactions with the system under evaluation.
This is called implicit feedback \cite{yue2010learning}.

Online evaluation is widely used because implicit feedback is generally cheap and easy to obtain, and it allows businesses to calculate useful metrics on end-users that interact directly with the ranking function (metrics like the rate of clicked documents).

This paper focuses on interleaved evaluation approaches, specifically, the Team-Draft Interleaving (TDI) method \cite{radlinski2008does}, applied to the comparison of two ranking functions.

Interleaving is an alternative to AB testing \cite{schuth2015predicting}.
It avoids the principal source of variance caused by the separation of the users in groups and the consequent necessary combination of the results \cite{schuth2015predicting}.
In AB testing one group is exposed to the control system and another group is exposed to the variant.
Each group sees just a single ranking function during the evaluation.

On the other hand, interleaved methods, when responding to a query, create a unique list with search results impartially picked from the two ranking functions.
This interleaved list is presented to the end-users transparently, so they don't know which result item is coming from which ranking function.

In previous works interleaved methods have been evaluated with a uniform distribution of queries \cite{hofmann2011probabilistic}.
Experimental setups have simulated users submitting a query by randomly sampling from a set of available queries (with replacement).

In real-world applications, the same query is executed multiple times by different users and by the same users in different sessions.
The number of collected implicit feedback is not uniformly distributed across the query set.
This paper aims to replicate and then reproduce some of the experiments from one of the most prominent surveys on interleaved methods \cite{hofmann2011probabilistic}, investigating the effect that different query distributions have on the accuracy of TDI, extending and generalizing its evaluation to different settings commonly found in production systems.

Specifically, three research questions arise:
\begin{itemize}
 \item $RQ1$: Is it possible to replicate the original paper experiments?
 \item $RQ2$: How does the original work generalize in the real-world scenario where queries have a long-tailed distribution?
 \item $RQ3$: Does applying statistical hypothesis testing improve the TDI evaluation accuracy in such a scenario?
\end{itemize}

Thanks to the insights collected during the reproducibility work we designed two novel methods that enrich the TDI scoring estimator with a preliminary statistical analysis: \textit{stat-pruning} and \textit{stat-weight}.
The idea is to weigh differently the contribution of each query to the final winner of the evaluation.
Such contribution should be proportional to the statistical significance of the observations collected for the query.

We present a set of experiments using a large set of explicitly labeled data and a framework, developed from scratch, to simulate the implicit feedback with user clicks under different conditions.
These experiments show some interesting perspectives on the original work replicability, confirm that TDI generalizes quite accurately to the considered real-world scenario, and validate the intuition that our statistical analysis-based methods can enrich TDI to bring better accuracy in identifying the best ranking function.

The concepts 'ranking function', 'ranking model', and 'ranker' are used interchangeably.

The paper is organized as follows:  \hyperlink{section.2}{Section 2} presents the related work.
\hyperlink{section.3}{Section 3} details the experimental setup, datasets, and runs used for replication and reproduction.
\hyperlink{section.4}{Section 4} introduces the theory behind the proposed improvements and describe our the \textit{stat-pruning} and \textit{stat-weight} implementation.
\hyperlink{section.5}{Section 5} discusses the experiments' runs and the obtained results.
The paper's conclusions and future directions are listed in \hyperlink{section.6}{Section 6}.

\section{Related Work}
Evaluation of Information Retrieval systems follows two approaches:
offline and online evaluation.

For the offline, the most commonly used is the Cranfield framework \cite{cleverdon1966factors}: an evaluation method based on explicit relevance judgments.
The experimental collection is composed of triples that represent: a document, a query, and the relevance of that document for that query.
The relevance judgments are provided by a team of trained experts and this is why this process is expensive. Collecting these judgments requires a lot of effort and there is the possibility that they do not reflect the same document relevance perceived by the common users.

Users' interactions are easier to obtain and come with a minimal cost.
Being performed directly by the end-users, they can be used to represent their intent closely, bypassing the domain experts' indirection.
Implicit feedback is a very promising approach but, as a drawback, it could be noisy and therefore requires some further elaborations.
One type of noise introduced is click position bias.
Users tend to click documents in the top positions of the search result list, independently of the relevance of the result.
Many papers focus on this topic and develop approaches to make interleaving fairer \cite{chapelle2012large,chuklin2013click,radlinski2008does,schuth2014multileaved}.
Implicit feedback is collected in real-time and it's at the base of the interleaving process.
Despite the fact that the most common method of online evaluation is still AB testing, interleaving is experiencing a growing interest in research.
This type of testing uses a smaller amount of traffic, with respect to AB testing, without losing accuracy in the obtained result \cite{chapelle2012large,kharitonov2013using,schuth2015predicting}.
Interleaving was introduced the first time by Joachims \cite{joachims2003evaluating,joachims2002optimizing} and from then, many other authors proposed their changes and improvements \cite{chapelle2012large,radlinski2013optimized,radlinski2008does,schuth2014multileaved,schuth2015probabilistic}.

Team Draft Interleaving (TDI from now on) is among the most successful and used interleaving approaches \cite{radlinski2008does} because of its simplicity of implementation and good accuracy.
It is based on the strategy that captains use to select their players in team matches.
TDI produces a fair distribution of ranking functions' elements in the final interleaved list.
It also showed to overcome issues of a previously implemented approach, Balanced interleaving, in determining the winning model \cite{chapelle2012large}.
\begin{figure}[h]
  \centering
  \includegraphics[width=\linewidth]{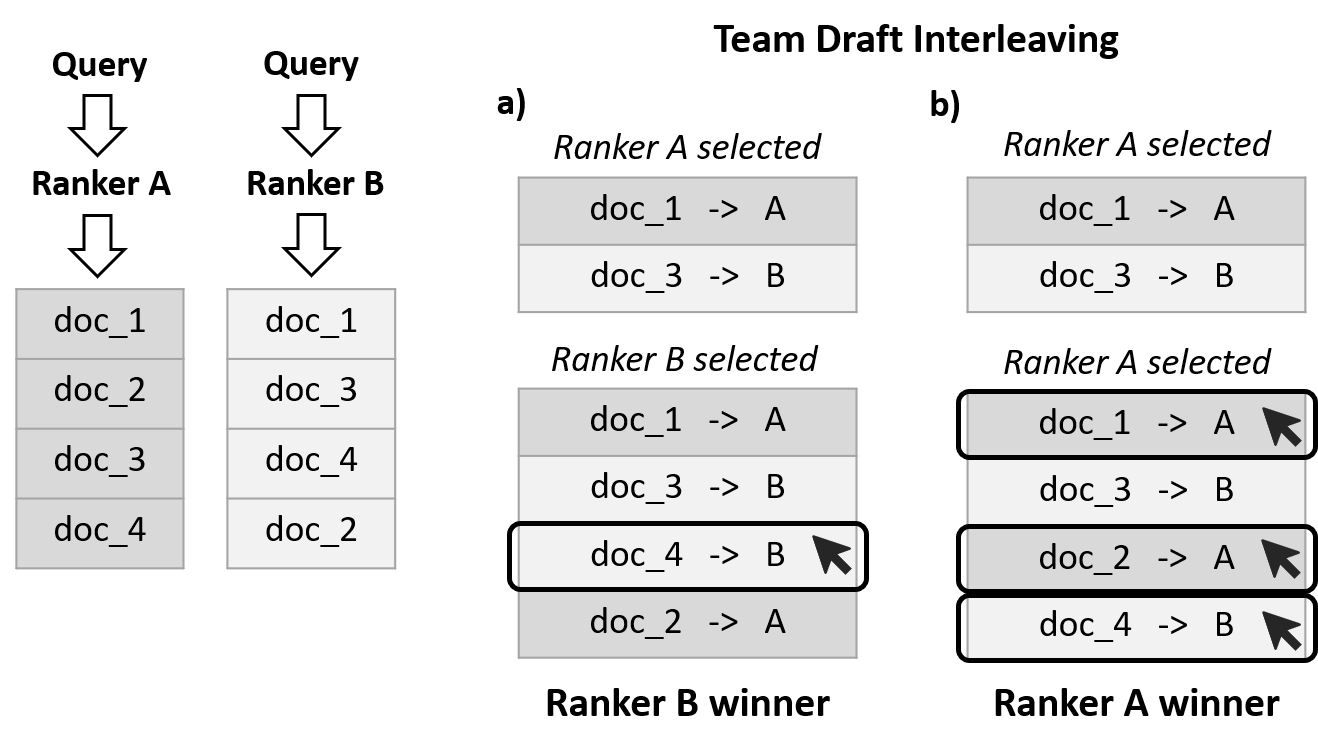}
  \caption{Example of the Team Draft Interleaving method. Two combinations (ABBA and ABAB are shown out of the four combinations possible)}
  \label{fig:TDI_figure}
\end{figure}

Team Draft Interleaving considers two ranking models: $rankerA$ and $rankerB$ (see \autoref{fig:TDI_figure}).
For a given query, each ranker returns its ranked list of documents $l_a=(a_1, a_2, ...)$ and $l_b=(b_1, b_2, ...)$.
The algorithm creates a unique ranked list $I=(i_1, i_2, ...)$.
This list is created by interleaving elements from the two lists $l_a$ and $l_b$ as described by Chapelle et al.\cite{chapelle2012large}.
Each element $i_j$ is labelled $TeamA$ if it is selected from $l_a$ and $TeamB$ if it is selected from $l_b$.

The list $I$ is returned to the user, who interacts with the search results of interest.
Let us consider clicks as a target interaction type for our analysis: given a click $c$, $c_j$ is the position of the clicked search result in the ranked list $I$.
Iterating on all clicks, the number of clicks collected by each ranker is computed as in Chapelle et al. \cite{chapelle2012large}:
\begin{equation*}
    \footnotesize
	h_a=|\{c_j: i_{c_j} \in TeamA\}|
\end{equation*}
\begin{equation*}
    \footnotesize
	h_b=|\{c_j: i_{c_j} \in TeamB\}|
\end{equation*}
If $h_a > h_b$ then $rankerA$ is the winner for that query; if $h_a < h_b$ then $rankerB$ is the winner; otherwise it is a tie.

To assess the overall winner between $rankerA$ and $rankerB$, the $\Delta_{AB}$ score is computed as \cite{chapelle2012large}:
\begin{equation}
    \label{eq:deltascore}
    \footnotesize
	\Delta_{AB} = \frac{wins(A) + \frac{1}{2}ties(A,B)}{wins(A) + wins(B) + ties(A,B)} - 0.5
\end{equation}
Where:
\begin{itemize}
\item $wins(A)$ is the number of queries in which $rankerA$ is the winner
\item $wins(B)$ is the number of queries in which $rankerB$ is the winner
\item $ties(A,B)$ is the number of queries in which the two rankers equalize
\end{itemize}
A $\Delta_{AB}$ score $<$ 0 means $rankerB$ is the overall winner, a $\Delta_{AB}$ score $=$ 0 means a tie, a $\Delta_{AB}$ score $>$ 0 means $rankerA$ is the overall winner.

Other types of interleaved methods are Document Constraint \cite{he2009evaluation}, Probabilistic Interleaving \cite{hofmann2011probabilistic} and Optimized Interleaving \cite{radlinski2013optimized}.

Document constraint infers preference relations between pairs of documents, estimated from their clicks and ranks.
The method compares the inferred constraints to the original result lists and assesses how many constraints are violated by each.
The list that violates fewer constraints is deemed better.
This method demonstrated to be more reliable than either balanced interleave or team draft on synthetic data, but it's more computationally expensive.

In probabilistic interleaving, both the choice of the model that contributes to the interleaving list and the document to put in the list, are selected based on probability. This approach shows higher reliability, efficiency, and robustness to the noise with respect to the others.

Optimized interleaving proposes to formulate interleaving as an optimization problem that is solved to obtain the interleaved lists that maximize the expected information gained from users' interactions.

To conclude this Section it's worth mentioning that a generalized form of the Team Draft Interleaving has been proposed \cite{kharitonov2015generalized} and that additional research has been performed by Hofmann et al. with a new interleaving approach that aims to reduce the bias related to the way results are presented to the users \cite{hofmann2012caption} and studies on the fidelity, soundness, and efficiency of interleaved methods \cite{hofmann2013fidelity}.

\section{Experiments}
\label{Experiments}
We present a set of experiments designed to answer our three research questions.
This paper aims to replicate and then reproduce under different scenarios experiment 1 from one of the most prominent surveys on interleaved methods \cite{hofmann2011probabilistic}.

In the first set of experiments, we address $RQ1$ with the same settings and data as the original work (uniform query distribution).
We compare a large number of ranker pairs over simulated user clicks to examine if the accuracy of the TDI method matches the published results from the original research.
We discuss the assumptions and details that we found to hold up, and the ones that could not be confirmed.
In addition we assess how the traditional TDI accuracy compares with two novel methods for $\Delta_{AB}$ score calculations (\textit{stat-pruning} and \textit{stat-weight}) under the uniform distribution conditions.
It has been chosen to reproduce this experiment with the TDI method for the simple implementation and the easier reproduction process. Even if probabilistic interleaving has been shown to be the best method, TDI maintains a good trade-off between performance and simplicity.

In the second set of experiments, we address $RQ2$ and $RQ3$ introducing experimental scenarios with a long-tailed query distribution.
We compare a large number of ranker pairs over different real-world query distributions extracted from anonymized query logs.
The aim is to examine how well the traditional TDI method generalizes in these real-world scenarios and how \textit{stat-pruning} and \textit{stat-weight} perform for comparison.

Finally, the last set of experiments introduces a realistic click model simulator to assess how well TDI, \textit{stat-pruning} and \textit{stat-weight} methods respond to noise.

The datasets used are detailed in \hyperlink{subsection.3.1}{3.1}.
The experimental setup is described in \hyperlink{subsection.3.2}{3.2} and the experiment runs are explained in \hyperlink{subsection.3.3}{3.3}.

\subsection{Datasets}
All experiments make use of the MSLR-WEB30k Microsoft learning to rank ($MSLR$) dataset\footnote[1]{download from \href{http://research.microsoft.com/en-us/projects/mslr/default.aspx. }{http://research.microsoft.com/en-us/projects/mslr/default.aspx.}}.
This dataset represents queries and documents by IDs.
It consists of feature vectors extracted from query-document pairs along with relevance judgment labels.
The relevance judgments are obtained explicitly from a retired labeling set of a commercial web search engine, which takes 5 values from 0 (irrelevant) to 4 (perfectly relevant).
The features describe aspects of the query-document pair widely used in the research community such as the length of a field of the document or the term frequency of the query terms in a field of the document.
In the data files, each row corresponds to a query-document pair.
The first element is the relevance label of the pair, the second is the query id, and the following elements are the features.

The experiments use the training set of fold 1.
This set contains $18,919$ queries, with an average of $119.96$ judged documents per query.
To define the ranking functions to compare in our experiments, we leverage the features provided for the documents of the dataset.
Specifically, from each feature, we define a ranker (identified by the feature id) that sorts the search results descending by the feature value.

The experiments involving the long-tailed distributions make use of an industrial dataset we called \textit{long-tail-1}. It consists of a list of query executions extracted from the query log of a commercial search engine over a period of time.
Each query is associated with the number of times is executed by different users.
The amount of users collected per query is capped to $1,000$. This threshold ensures maintaining the long-tailed distribution while keeping a sustainable computational cost.

\begin{figure}[h]
  \label{fig:long-tail-1}
  \centering
  \includegraphics[width=\linewidth]{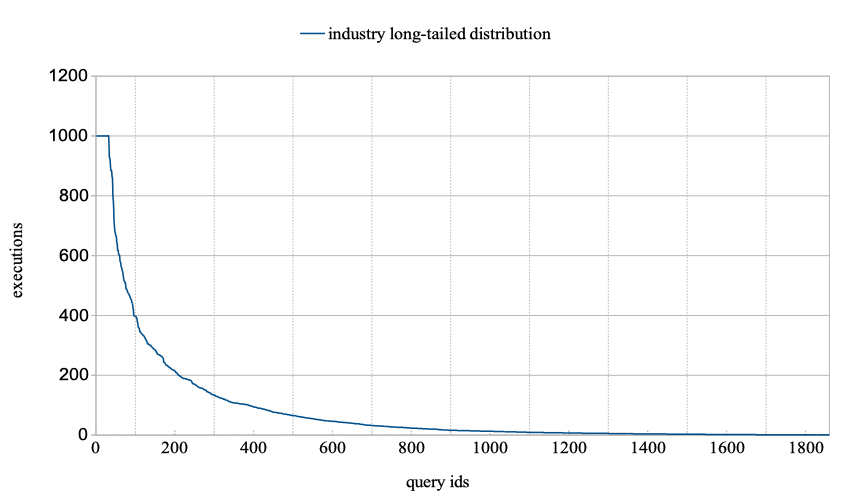}
  \caption{Total unique queries: $1\,861$, Total executions:$156\,550$}
  \label{fig:long_tail}
\end{figure}

From this dataset, we derive the long tail in \autoref{fig:long_tail}.

\subsection{Experimental Setup}
To replicate the original experiments we designed and developed a system that simulates a search engine with users submitting queries and interacting with the results (clicks).
The code references are in the appendix.

The experiments are designed to evaluate the interleaved methods’ ability to establish the better of two ranking functions based on (simulated) user clicks.

When a query is submitted to the simulated search engine it returns the pre-calculated list of matching search results with explicit relevance judgments from the dataset $MSLR$.
The result set is ranked by a TDI interleaved evaluation of two ranking functions.

Each experiment run repeats a number of simulations proportional to the number $r$ of different ranking functions we want to include in the evaluation.
Each ranking function is identified by an incremental id that is aligned with the id of the feature of reference.
When an experiment evaluates $r$ ranking functions, it evaluates the first $r$, ordered by ascending id.
Specifically, given a set of ranking functions, the number of simulations $s$ in the run is the number of unique pairs in the set, where the pairs are subject to the commutative property (AB = BA).
The user interactions are simulated using a query distribution and a click model.

The system simulates a user submitting a query from the set of available queries in the distribution (in long-tailed distributions each query is submitted multiple times).
The search engine responds with an interleaved result list that is presented to the user.
The simulation models the assumption that more relevant documents are more likely to be clicked.
The user clicks are randomly generated following the probability distribution that the click model assigns to the relevance judgments provided.

Once the simulation completes, the ranking function preference of each click collected is evaluated and the $\Delta_{AB}$ score is computed to establish the winner by the interleaving evaluation chosen.
The ground truth winner is calculated as the ranking function with the best average Normalised Discounted Cumulative Gain (NDCG from now on) \cite{jarvelin2017ir,jarvelin2002cumulated} on the considered query set.
The NDCG calculation is based on the explicit relevance judgments provided with the dataset $MSLR$.
Depending on the run, to determine the NDCG we use the complete search results list for a query or a top-k (cut-off at k).
The winning ranker identified by the $\Delta_{AB}$ score is compared to the ground truth winner for the pair, when they match we have a correct guess.
To assess the accuracy of the interleaved evaluation method we count the number of correct guesses over the total number of simulations $s$ in the run showing at least one click.

Below we describe the query distributions, the click models, and the NDCG we used in our experiments.

\paragraph{Query Distributions}
The query distribution in input to the simulation establishes the number of queries that the user submits to the system.
We use two types of query distributions in our experiments: uniform and long-tailed.

In the uniform query distribution, each unique query is executed a constant number of times.
When considering $q$ queries in a run, we refer to the first $q$ query ids, in the same order as they appear in the dataset $MSLR$ rows.

\begin{figure}[h]
 \centering
 \includegraphics[width=\linewidth]{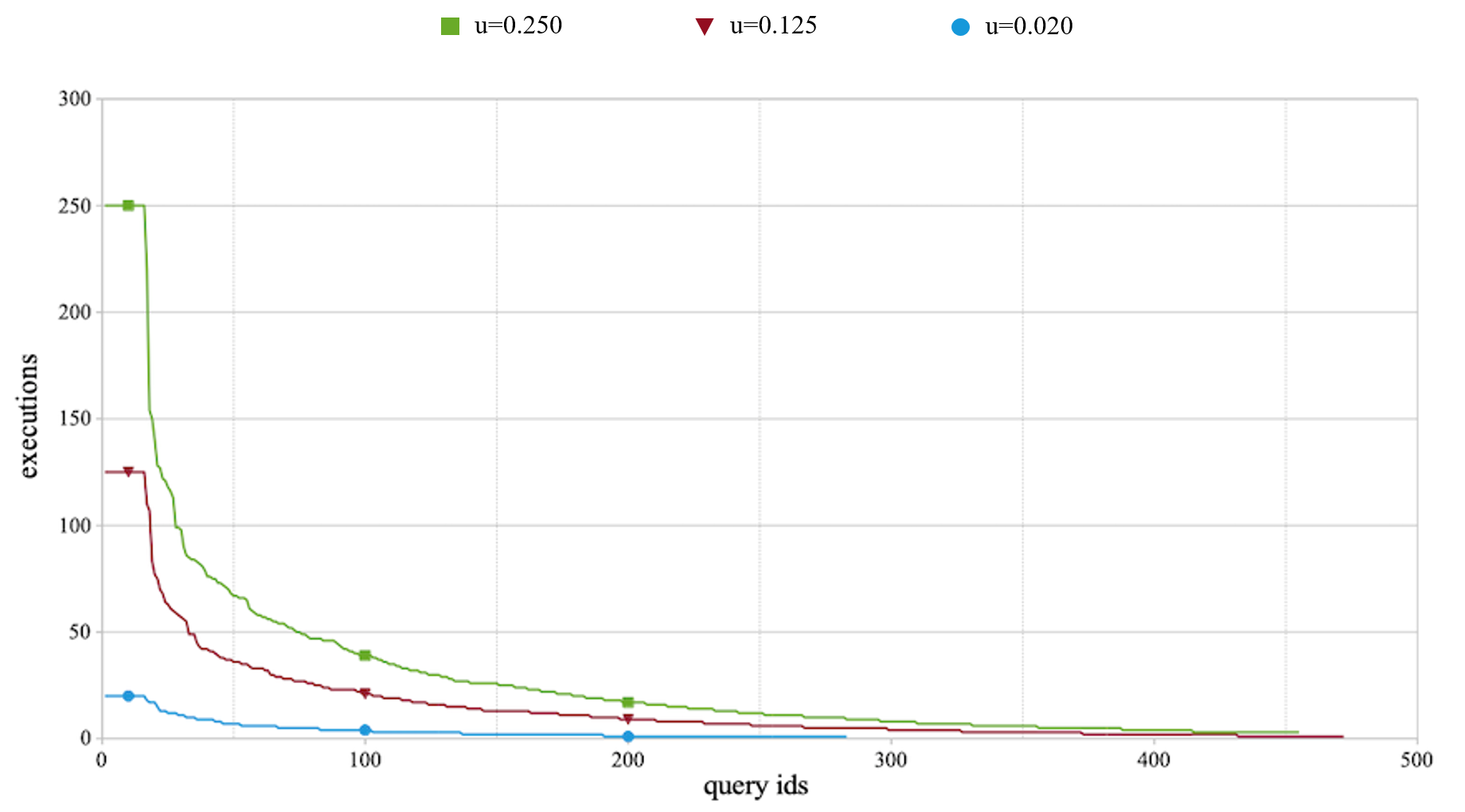}
 \caption{Long-tailed query distributions used in the experiments}
 \label{fig:long_tail_variants}
\end{figure}

In the long-tailed query distribution, each query is executed a variable number of times.
Starting from the \textit{long-tail-1} distribution from the industry, we scaled down the number of queries and their executions by a factor \begin{math} u\leq1\end{math} (see \autoref{fig:long_tail_variants}).
The first reason for that is to experiment how the evaluation methods perform with different instances of realistic long-tailed distributions.
The second reason is to act within the computational limits of our experimental setup.
For each ranker pair to interleave, we simulate a number of query executions taken from the query distribution.
The closer $u$ gets to 1, the higher the total number of executions (see \autoref{tab:long-t}).
\begin{table}[h]
  \centering
  \caption{Scaling the query executions}
  \label{tab:long-t}
  \begin{tabular}{c||cc}
    \toprule
    \textit{u}& \textit{unique queries}& \textit{total executions}\\
    \hline
    0.020& 283& 1247\\
    0.125& 472& 7681\\
    0.250& 455& 14449\\
    \bottomrule
\end{tabular}
\end{table}
Higher the number of executions higher the cost in terms of memory and time consumption.
Running the full simulation, on all the rankers, with the original long-tailed distribution would have required too much time and memory in our setup.

\paragraph{Click Models} The click model in input to the simulation establishes the probability of a search result to be clicked given its explicit relevance label (ground truth).
Using different click models makes it possible to study the evaluation methods at different levels of noise in user clicks.

The click model simulates user interactions according to the Dependent Click Model (DCM) \cite{guo2009tailoring,guo2009efficient}, an extension of the cascade model \cite{craswell2008experimental}.

According to this model, users scan result lists from the top-ranking result to the last.
For each document they see, they decide to click it or not, depending on its perceived relevance (e.g., based on the title, thumbnail, and content snippets).
After clicking on a document, users assess the full relevance of the search result and if the information need is satisfied then they stop scanning the result list.
Otherwise, they progress.
\begin{table}[h]
  \centering
  \caption{Click models used in the experiments \cite{hofmann2011probabilistic}}
  \label{tab:click_models}
  \begin{tabular}{c||ccccc}
    \toprule
    \multicolumn{6}{c}{\textit{perfect model}} \\
    \hline
    \hline
    $R(d)$& 0& 1& 2& 3& 4\\
    $P(c\mid R(d))$& 0.00& 0.20& 0.40& 0.80& 1.00\\
    $P(s\mid R(d))$& 0.00& 0.00& 0.00& 0.00& 0.00\\
    \hline
    \multicolumn{6}{c}{\textit{realistic model}} \\
    \hline
    \hline
    $R(d)$& 0& 1& 2& 3& 4\\
    $P(c\mid R(d))$& 0.05& 0.10& 0.20& 0.40& 0.80\\
    $P(s\mid R(d))$& 0.00& 0.20& 0.40& 0.60& 0.80\\
  \bottomrule
\end{tabular}
\end{table}

For each document viewed by the user, $P(c\mid R(d))$ is the probability that a click is performed given the relevance label of the examined document $R(d)$.
$P(s\mid R(d))$ is the probability that a user stops scanning the result list because it has satisfied the information need after clicking a document with the relevance label $R(d)$.

We use the two-click models proposed by the original research \cite{hofmann2011probabilistic}.
The realistic click model presents a noisier click behavior.
Furthermore, the overall expected clicks quantity is lower for the realistic click model because the click probabilities are consistently lower and the stop probabilities are higher.
For these reasons, the realistic click model increases the difficulty for the interleaved methods to correctly guess the best ranking function.
\autoref{tab:click_models} lists the probabilities defined for the two models.

\paragraph{NDCG}
The NDCG metrics we used in our experiments are the complete NDCG (the complete search results list for a query) and the NDCG@10 (cut-off at 10).

Using NDCG@10 is quite common in the industry as many search engines show 10 documents on their first page and it aligns with the decision of having the simulated users click only the top-10 results.
When comparing the complete NDCG with NDCG@10 we noticed that the average difference between the pair of rankers to evaluate is smaller, making it more difficult for the interleaved methods to correctly guess the best ranker (see \autoref{tab:ndcg} and \autoref{tab:ndcg2}).
\begin{table}[h]
  \centering
  \caption{complete NDCG and NDCG@10 for the first 10 rankers, averaged over the 1000 queries}
  \label{tab:ndcg}
  \begin{tabular}{c||cc}
    \toprule
    \textit{ranker}& \textit{NDCG}& \textit{NDCG@10}\\
    \hline
    1& $0.550$& $0.195$\\
    2& $0.578$& $0.260$\\
    3& $0.603$& $0.284$\\
    4& $0.585$& $0.261$\\
    5& $0.550$& $0.192$\\
    6& $0.550$& $0.195$\\
    7& $0.578$& $0.260$\\
    8& $0.603$& $0.284$\\
    9& $0.585$& $0.261$\\
    10& $0.550$& $0.192$\\
    \bottomrule
\end{tabular}
\end{table}

\begin{table}
  \centering
  \caption{average difference in NDCG over the $9\,180$ pairs of rankers}
  \label{tab:ndcg2}
  \begin{tabular}{c||cc}
    \toprule
    \textit{NDCG}& \textit{NDCG@10}\\
    \hline
    $0.026$& $0.046$\\
    \bottomrule
\end{tabular}
\end{table}

\subsection{Runs}
We divided the runs into four groups: replication, uniform query distribution, long-tailed query distribution, and realistic click model.
A seed is set at the beginning of each run so that it is possible to reproduce all the random choices of the experiment reliably in repeated executions.
Using different seeds will result in slightly different results.
An interesting future study could be to execute each run multiple times with different seeds and explore such differences from a statistical perspective.

\paragraph{Replication}
The scope of this set of runs is to replicate the \textit{experiment 1} from the original research and answer $RQ1$.
As a baseline, to reduce the computational stress we evaluate the Team Draft Interleaving method only.
We define a ranker for each of the $136$ individual features provided with the $MSLR$ dataset.
We exhaustively compare all $9,180$ distinct pairs derived from the $136$ rankers.
The query distribution used is uniform and for each ranker pair, the user submits $1,000$ queries.
The query set consists of the first distinct $1,000$ queries as occurring in the $MSLR$ dataset rows.
The click model used is the \textit{perfect model}.
The results report the percentage of pairs for which the TDI method correctly identified the better ranker.
\hfill\break
\begin{itemize}
    \item \textbf{Run 1}: exactly replicates the original experiment, users click on the top-$10$ results for each query, to determine NDCG for the ground truth we use the complete set of documents provided with the dataset $MSLR$, i.e., no cut-off is used.
\end{itemize}
\hfill\break
We observed some inconsistencies with the original work results so we added to this group two additional runs:
\hfill\break
\begin{itemize}
    \item \textbf{Run 2}: users click on the complete list of search results for each query. To determine NDCG for the ground truth we use the complete set of documents provided with the dataset $MSLR$, i.e., no cut-off is used.
    \item \textbf{Run 3}: users click on the top-$10$ results for each query.
    To determine NDCG for the ground truth we calculate NDCG@10.
\end{itemize}

\paragraph{Uniform query distribution}
The scope of this set of runs is to evaluate how the \textit{stat-pruning} and \textit{stat-weight} methods compare with the TDI baseline.
We exhaustively compare all $9,180$ distinct pairs derived from the $136$ rankers.
The query distribution used is uniform, each run uses a different number $q$ of queries.
The query set consists of the first distinct $q$ queries as occurring in the $MSLR$ dataset rows.
The click model used is the \textit{perfect model}.
Users click on the top-$10$ results for each query.
Unless stated otherwise, to determine NDCG for the ground truth we calculate NDCG@10.
The results compare the percentage of pairs for which the TDI, \textit{stat-pruning} and \textit{stat-weight} methods correctly identified the better ranker.
\hfill\break
\begin{itemize}
    \item \textbf{Run 4}: the query set consists of the first distinct $1,000$ queries as occurring in the $MSLR$ dataset. Each query is executed once.
    \item \textbf{Run 5}: the query set consists of the first distinct $100$ queries as occurring in the $MSLR$ dataset. Each query is executed once.
    \item \textbf{Run 6}: the query set consists of the first distinct $100$ queries as occurring in the $MSLR$ dataset. Each query is executed $10$ times.
    \item \textbf{Run 7}: the query set consists of the first distinct $100$ queries as occurring in the $MSLR$ dataset. Each query is executed $10$ times.
    To determine NDCG for the ground truth we use the complete set of documents provided with the dataset $MSLR$, i.e., no cut-off is used.
\end{itemize}

\paragraph{Long-tailed query distribution}
The scope of this set of runs is to evaluate how the \textit{stat-pruning} and \textit{stat-weight} methods compare with the TDI baseline over long-tailed query distributions and answer $RQ2$ and $RQ3$.
The click model used is the \textit{perfect model}.
Each run uses a different long-tailed query distribution, see \autoref{fig:long_tail} and \autoref{tab:long-t} for reference.
Users click on the top-$10$ results for each query.
Unless stated otherwise, to determine NDCG for the ground truth we calculate NDCG@10.
The results compares the percentage of pairs for which the TDI, \textit{stat-pruning} and \textit{stat-weight} methods correctly identified the better ranker.
\hfill\break
\begin{itemize}
    \item \textbf{Run 8}: the query set consists of $283$ unique queries repeated following the long-tailed distribution with $u$ = $0.020$.
    We exhaustively compare all $9,180$ distinct pairs derived from the $136$ rankers.
    \item \textbf{Run 9}: the query set consists of $472$ unique queries repeated following the long-tailed distribution with $u$ = $0.125$.
    We exhaustively compare all $9,180$ distinct pairs derived from the $136$ rankers.
    \item \textbf{Run 10}: the query set consists of $455$ unique queries repeated following the long-tailed distribution with $u$ = $0.250$.
    We exhaustively compare all $2,415$ distinct pairs derived from the first $70$ rankers.
    \item \textbf{Run 11}: the query set consists of $283$ unique queries repeated following the long-tailed distribution with $u$ = $0.020$.
    We exhaustively compare all $9,180$ distinct pairs derived from the $136$ rankers.
    To determine NDCG for the ground truth we use the complete set of documents provided with the dataset $MSLR$, i.e., no cut-off is used.
\end{itemize}

\paragraph{Realistic click model}
The scope of this set of runs is to evaluate how the \textit{stat-pruning} and \textit{stat-weight} methods compare with the TDI baseline over long-tailed query distributions with noisier clicks.
We exhaustively compare all $9,180$ distinct pairs derived from the $136$ rankers.
The click model used is the \textit{realistic model}.
The query distribution used is long-tailed.
The query set consists of $283$ unique queries repeated following the long-tailed distribution with $u$ = $0.020$, see \autoref{fig:long_tail} and \autoref{tab:long-t} for reference.
Users click on the top-$10$ results for each query.
The results compares the percentage of pairs for which the TDI, \textit{stat-pruning} and \textit{stat-weight} methods correctly identified the better ranker.
\hfill\break
\begin{itemize}
    \item \textbf{Run 12}: to determine NDCG for the ground truth we calculate NDCG@10.
    \item \textbf{Run 13}: to determine NDCG for the ground truth we use the complete set of documents provided with the dataset $MSLR$, i.e., no cut-off is used.
\end{itemize}

\section{Improving the Overall Winner Decision}
The research questions $RQ1$ and $RQ2$ are addressed by the experiments in Sections \ref{Experiments} and \ref{Results}, here we want to focus on $RQ3$ which requires a deeper analysis.

In TDI, for each query in the dataset, a winner between two ranking functions is estimated.
The estimation is based on the number of interactions (clicks for example) that prefer a ranking function or the other.
In the interleaved result list, an interaction with a document that was picked from the ranked list A (the original ranked list produced by $rankerA$), shows a preference for $rankerA$ and vice versa for $rankerB$.

The reliability of each winner is not assessed.
All the winners (i.e., all the queries) are considered equal when aggregating the results to establish the overall winning ranker (\autoref{eq:deltascore}).

This may include preferences that are obtained with few clicks or preferences that are not strong enough given the number of clicks collected.

To mitigate this problem, this paper proposes two variations for the $\Delta_{AB}$ score: \textit{stat-pruning} and \textit{stat-weight}.
They rely on an additional phase in the TDI evaluation, which assigns to each query a credit inversely proportional to the probability of obtaining by chance at least the same number of clicks, assuming the two rankers are equivalent.
This credit per query affects the overall winner calculation. It should reduce the impact of queries with a weak winning ranker leading to a more accurate overall winner estimation.

\subsection{Statistical hypothesis testing}
While performing our reproducibility research we observed two problems:
\begin{itemize}
    \item some queries have many interactions, but a very weak preference for the winning ranker
    \item some queries have  a decent preference for the winning ranker but few interactions (the long tail)
\end{itemize}

In classic TDI, each query has the same weight when calculating the $\Delta_{AB}$ score.
The overall winner decision may be polluted by the aforementioned queries.

Previous works have explored the possibility of assigning a different credit to each click \cite{radlinski2013optimized}.
The approach we suggest is to assign a different credit to each query.

The idea is to exploit statistical hypothesis testing to estimate if the observations for a query are reliable and to what extent.
This additional phase is executed after the computation of $h_a$ and $h_b$, and before the computation of the $\Delta_{AB}$ score.

The theory behind our approach is statistical hypothesis testing \cite{sirkin2006statistics}.
A statistical test verifies or contradicts a null hypothesis based on the collected samples.
A result has statistical significance when it is very unlikely to have occurred given the null hypothesis \cite{myers2010developing}.

The \textit{p-value} of an observed result is the probability of obtaining a result at least as extreme, given that the null hypothesis is true.
The result is statistically significant, by the standards of the study, when
\begin{equation*}
    \footnotesize
    \textit{p-value} <= \alpha
\end{equation*}

Such a scenario leads to the rejection of the null hypothesis and acceptance of the alternate hypothesis.
The significance level, denoted by $\alpha$ is assigned at the beginning of a study \cite{dalgaard2008power}.

Accordingly for this test, we need a null hypothesis, a p-value, and a significance level.

Our null hypothesis is that the two ranking functions we are comparing are equivalent and have the same chance to win in a query i.e. the probability of both ranking functions winning is 0.5.

For each query:
\begin{itemize}
    \item $n$ is the total number of clicks collected
    \item the winning ranker is the ranker that collected more clicks
    \item $k$ is the clicks collected by the winning ranker.
    \item $p$ is $0.5$ (null hypothesis).
\end{itemize}

Given we are limiting our evaluation to two ranking functions:

\begin{equation*}
  k\geq\frac{n}{2}
\end{equation*}

When
\begin{math}
  k = \frac{n}{2}
\end{math},
there is a draw, the query doesn't show any preference for the ranking function since each ranker collects the same amount of clicks.

The \textit{p-value} is calculated through a binomial distribution as the probability of obtaining exactly that number of clicks $k$ assuming the null hypothesis is true:

\begin{equation}
    \label{eq:tie}
    \small
    \begin{aligned}
    P(X = k) &= {{n\choose k}p^k(1-p)^{n-k}}
    \end{aligned}
\end{equation}

When
\begin{math}
  k > \frac{n}{2}
\end{math},
the query shows a preference for a ranking function.
We are testing whether the clicks are biased towards the winning ranking function, so a single-tailed test is used.

The \textit{p-value} is calculated through a binomial distribution as the probability, for the winning model, to obtain at least that number of clicks $k$ assuming the null hypothesis is true:

\begin{equation}
    \label{eq:win}
    \small
    \begin{aligned}
    P(X\geq k) &= 1 - P(X < k) \\
    & =  1 - \sum_{i=0}^{k-1}{{n\choose i}p^i(1-p)^{n-i}}
    \end{aligned}
\end{equation}

\subsection{Stat-pruning}
The  first approach we designed is the simplest and most aggressive:
the statistical significance of each query is determined by comparing the \textit{p-value} with a significance level $\alpha$ = 0.05. This is the standard threshold used in most statistical tests.

If the \textit{p-value} is below the threshold, the result is considered significant.
The queries not reaching significance are discarded from the $\Delta_{AB}$ score calculation.

The downside of this approach is that is strictly coupled to the significance level $\alpha$ hyper-parameter.
The setting of this parameter may not be simple to decide, other works explore this aspect \cite{queen2002experimental,sproull2002handbook}.

\subsection{Stat-weight}
Let's present for simplicity the original $\Delta_{AB}$ score formula again (\autoref{eq:deltascore}):

\begin{equation*}
    \small
	\Delta_{AB} = \frac{wins(A) + \frac{1}{2}ties(A,B)}{wins(A) + wins(B) + ties(A,B)} - 0.5
\end{equation*}

The credit associated with each win or tie is a constant 1.

The idea is to assign a different credit to each win and tie.
This credit is the estimated probability of the win/tie to have happened not by chance.

\begin{equation*}
    \small
    credit(q_x) = 1 - \textit{p-value}(q_x)
\end{equation*}

The \textit{p-value} for a query $q_x$ that presents a tie is calculated with the \autoref{eq:tie}.

The \textit{p-value} for a query $q_x$ that presents a win is calculated with the \autoref{eq:win} and it is normalised with a min-max normalization (\begin{math}
  min=0
\end{math} and \begin{math}
  max=0.5
\end{math}) to be between 0 and 1.

The proposed updates to the $\Delta_{AB}$ score formula are the following:

\begin{equation*}
    \small
    wins(A) \Rightarrow \sum_{a=0}^{wins(A)}{credit(q_a)}
\end{equation*}

\begin{equation*}
    \small
    wins(B) \Rightarrow \sum_{b=0}^{wins(B)}{credit(q_b)}
\end{equation*}

\begin{equation*}
    \small
    ties(A,B) \Rightarrow \sum_{t=0}^{ties(A,B)}{credit(q_t)}
\end{equation*}

$q_a$ is in the query set showing a preference for the $rankerA$.

$q_b$ is in the query set showing a preference for the $rankerB$.

$q_t$ belongs to the query set showing a tie.

\section{Results and Analysis}
\label{Results}
In this Section, we present and discuss the results of our experiments.

\subsection{Replication}
\begin{table}[h]
  \centering
  \caption{Query distribution: uniform $1,000$ queries, click model: \textit{perfect model}, 136 rankers ($9,180$ pairs) }
  \label{tab:run1}
  \begin{tabular}{c|cc||cc}
    \toprule
    \textit{id}& \textit{NDCG}& \textit{clicks}& \textit{accuracy}& \textit{original-accuracy}\\
    \hline
    1& complete& top-10& 0.852& $0.898$ \\
    2& complete& complete& 0.825& $0.898$\\
    3& top-10& top-10& \textbf{0.902}& $0.898$\\
    \bottomrule
\end{tabular}
\end{table}
\textit{run-1} follows the same experimental setup, dataset, and parameters from the original research, but it fails to replicate the originally recorded accuracy of TDI.

Also, the average ground truth NDCGs calculated for the rankers, don't align with the ones reported by the original paper (\autoref{tab:run1ndcg}):

\begin{table}[h]
  \centering
  \caption{complete NDCG, averaged over the 1000 queries}
  \label{tab:run1ndcg}
  \begin{tabular}{c||cc}
    \toprule
    \textit{ranker}& \textit{NDCG}& \textit{original-paper NDCG}\\
    \hline
    1& $0.550$& $0.231$\\
    14& $0.536$& $0.201$\\
    64& $0.600$& $0.301$\\
    77& $0.570$& $0.262$\\
    84& $0.574$& $0.256$\\
    96& $0.549$& $0.219$\\
    97& $0.564$& $0.303$\\
    106& $0.606$& $0.253$\\
    108& $0.614$& $0.306$\\
    134& $0.614$& $0.341$\\
    \bottomrule
\end{tabular}
\end{table}
We thought that the difference could have been caused by a dis alignment between the published paper and the original experiments NDCG and clicks generation parameters used at the time.
For these reasons we executed two additional runs, trying to explain the possible causes of this failed replication.
The closer we got to the original recorded accuracy is with \textit{run-3}, but not exactly the same (see \autoref{tab:run1}).
Also, the monitored average NDCG@10 over the $1,000$ queries are closer but not exactly matching the original work ones (\autoref{tab:run1ndcg2}).
\begin{table}[h]
  \centering
  \caption{NDCG@10, averaged over the 1,000 queries}
  \label{tab:run1ndcg2}
  \begin{tabular}{c||cc}
    \toprule
    \textit{ranker}& \textit{NDCG@10}& \textit{original-paper NDCG}\\
    \hline
    1& $0.195$& $0.231$\\
    14& $0.179$& $0.201$\\
    64& $0.294$& $0.301$\\
    77& $0.246$& $0.262$\\
    84& $0.239$& $0.256$\\
    96& $0.194$& $0.219$\\
    97& $0.234$& $0.303$\\
    106& $0.295$& $0.253$\\
    108& $0.306$& $0.306$\\
    134& $0.333$& $0.341$\\
    \bottomrule
\end{tabular}
\end{table}
The random seed that drives the TDI interleaved-lists generation and the clicks can have a part, but it can't affect the ground truth NDCG scores.

After discussing with the authors of the paper, we could ascertain that the NDCG formula used is the same as ours. However, we weren't able to check the input parameters since we couldn't have access to the original paper code.
Our best guesses are therefore the following:
\hfill\break
\begin{itemize}
    \item \textbf{NDCG}: the published paper clearly specifies it is the complete NDCG, but the original experiments maybe initially used it and then were updated to use the NDCG@10 for the final reports.
    This could explain why NDCG@10 scores are much closer to the reported ones.
    \item \textbf{Queries}: the query set used is not the same as our runs i.e., not the first distinct $1,000$ queries as occurring in the $MSLR$ dataset rows.
    This could explain why NDCG@10 scores are closer but not exactly the same as the reported ones.
\end{itemize}
\hfill\break

\subsection{Uniform Query Distribution}
\begin{table}[h]
  \centering
  \caption{NDCG: top-10, clicks: top-10, click model: \textit{perfect model}, 136 rankers ($9,180$ pairs) }
  \label{tab:run4}
  \begin{tabular}{c|cc||ccc}
    \multicolumn{3}{c}{}& \multicolumn{3}{c}{\textit{accuracy}} \\
    \toprule
    \textit{id}& \textit{queries}& \textit{users}& \textit{TDI}& \textit{stat-pruning}& \textit{stat-weight}\\
    \hline
    4& $1\,000$& 1& \textbf{0.902}& \textit{N/A}& $0.886$\\
    5& $100$& 1& \textbf{0.812}& \textit{N/A}& $0.790$\\
    6& $100$& 10& $0.857$& $0.853$& \textbf{0.883}\\
    \bottomrule
\end{tabular}
\end{table}

From \autoref{tab:run4}, \textit{run-4} and \textit{run-5} show a better accuracy for the original TDI method with a uniform distribution of $100$ or $1,000$ queries if each query is executed once.
In these scenarios there are very few clicks per query, the accuracy for \textit{stat-pruning} is not available as it removes aggressively all the queries as deemed not significant.
\textit{stat-weight} doesn't shine as well: it has too few clicks per query to work on.
This is a little unrealistic for real-world use cases, even if uniform.
So \textit{run-6} explores what happens if the distribution is uniform and each query is executed 10 times.

In this scenario, we start to see the benefits of the \textit{stat-weight} approach with a $2.6\%$ increase in accuracy (it correctly guessed 239 additional pairs) in comparison to the classic TDI.

Comparing \textit{run-5} and \textit{run-6} we notice that by increasing the number of users running the queries uniformly, all the methods improve their accuracy and converge more quickly.
This is expected as we get more clicks per query and it's interesting to notice that \textit{stat-weight} is able to better handle the additional interactions discerning where they are reliable or not to identify the best ranker.

\textit{stat-weight} was originally designed for long-tailed distributions, but the take away from this set of runs is that it is quicker to converge and can perform better than TDI also in uniform distributions, with the caveat that queries are repeated more than once by the users.

\textit{stat-pruning} showed to be generally too aggressive and it's the worst in terms of accuracy (using $\alpha$ = $0.05$).

\begin{table}[h]
  \centering
  \caption{NDCG: complete, clicks: complete, click model: \textit{perfect model}, 136 rankers ($9,180$ pairs) }
  \label{tab:run7}
  \begin{tabular}{c|cc||ccc}
  \multicolumn{3}{c}{}& \multicolumn{3}{c}{\textit{accuracy}} \\
    \toprule
    \textit{id}& \textit{queries}& \textit{users}& \textit{TDI}& \textit{stat-pruning}& \textit{stat-weight}\\
    \hline
    7& $100$& 10& $0.828$& $0.839$& \textbf{0.857}\\
    \bottomrule
\end{tabular}
\end{table}
From \autoref{tab:run7}, \textit{run-7} makes the task more difficult as the complete NDCG presents less difference between the rankers, so it's more challenging for the interleaving methods to guess correctly.
See \autoref{tab:ndcg} and \autoref{tab:ndcg2} for NDCG comparisons.

\textit{stat-weight} demonstrated to be more sensitive in this difficult scenario identifying correctly 267 additional pairs.

\subsection{Long-tailed Query Distribution}
\begin{table}
  \centering
  \caption{NDCG: top-10, clicks: top-10, click model: \textit{perfect model}}
  \label{tab:run8}
  \begin{tabular}{c|cc||ccc}
  \multicolumn{3}{c}{}& \multicolumn{3}{c}{\textit{accuracy}} \\
    \toprule
    \textit{id}& \textit{u}& \textit{rankers}& \textit{TDI}& \textit{stat-pruning}& \textit{stat-weight}\\
    \hline
    8& 0.020& 136& 0.880& 0.860& \textbf{0.897}\\
    9& 0.125& 136& 0.892& 0.900& \textbf{0.904}\\
    10& 0.250& 70& 0.904& 0.910& \textbf{0.911}\\
    \bottomrule
\end{tabular}
\end{table}
From \autoref{tab:run8}, \textit{run-8}, \textit{run-9} and \textit{run-10} explore different long-tailed distributions.
Due to computational limits, we had to restrict the number of rankers the closer we were getting to the original \textit{long-tailed-1} distribution.
For this reason, it's not fair to compare the executions against each other and each must be observed independently.

The steepest the long-tail, the better \textit{stat-pruning} performs.
This is expected as we assume the long part of the tail to add uncertainty for TDI, an uncertainty that is cut by the \textit{stat-pruning} and mitigated by the \textit{stat-weight} approach.
This confirms the intuition that statistical hypothesis testing improves the classic TDI $\Delta_{AB}$ score accuracy in the long-tailed scenario.

\begin{table}[h]
  \centering
  \caption{NDCG: complete, clicks: top-10, click model: \textit{perfect model}, 136 rankers (9180 pairs) }
  \label{tab:run11}
  \begin{tabular}{c|cc||ccc}
  \multicolumn{3}{c}{}& \multicolumn{3}{c}{\textit{accuracy}} \\
    \toprule
    \textit{id}& \textit{u}& \textit{rankers}& \textit{TDI}& \textit{stat-pruning}& \textit{stat-weight}\\
    \hline
    11& 0.020& 136& 0.827& 0.817& \textbf{0.837}\\
    \bottomrule
\end{tabular}
\end{table}
From \autoref{tab:run11}, \textit{run-11} explores again the harder problem of closer rankers with the complete NDCG.
We can see that \textit{stat-weight} confirms its sensitivity and it is able to identify correctly 91 additional pairs also in the long-tailed scenario.

\subsection{Realistic Click Model}
\begin{table}
  \centering
  \caption{Query distribution: long-tailed $u$=0.020, clicks: top-10, click model: \textit{realistic model}, 136 rankers ($9,180$ pairs) }
  \label{tab:run12}
  \begin{tabular}{c|c||ccc}
  \multicolumn{2}{c}{}& \multicolumn{3}{c}{\textit{accuracy}} \\
    \toprule
    \textit{id}& \textit{NDCG}& \textit{TDI}& \textit{stat-pruning}& \textit{stat-weight}\\
    \hline
    12& top-10& 0.818& 0.708& \textbf{0.833}\\
    13& complete& 0.782& 0.693& \textbf{0.795}\\
    \bottomrule
\end{tabular}
\end{table}
From \autoref{tab:run12}, \textit{run-12} introduces an additional challenge with the \textit{realistic model} that produces noisier and fewer clicks, so it's more difficult for the interleaving methods to guess correctly.
\textit{stat-weight} demonstrated to be robust to noise with a $1.5\%$ increase (137 additional pairs) in comparison to the classic TDI.
Finally, \textit{run-13} tests the methods under an even more difficult situation with the complete NDCG.
The overall scores across the three methods are smaller, but the \textit{stat-weight} keeps consistently the lead.

\section{Conclusions And Future Directions}

$RQ1$ has not been satisfied.

Replicating the original research turns to be challenging from many angles: it was easy to align with the datasets but it required a substantial amount of work to figure out the exact parameters used in the original runs and design and develop from scratch the experiment code to cover all the necessary scenarios.

Unfortunately, it was not possible to exactly replicate the reported accuracy for TDI due to missing information and code unavailability. After a discussion with the authors of the paper, two hypotheses have been made: NDCG reported as a complete NDCG was instead using a cut-off; the query set is different from the one we used (which 1000 queries?).

$RQ2$ has been satisfied.
We verified that it is possible to generalize the original TDI evaluation to long-tailed query distributions with good accuracy.

$RQ3$ has been satisfied.
The reproducibility work brought many interesting insights and we developed a new set of statistical-based $\Delta_{AB}$ score methods: \textit{stat-pruning} and \textit{stat-weight}.
Applying statistical hypothesis testing to assign a credit score to each of the queries in the evaluations has shown to be promising because adapts quite well to various real-world scenarios and doesn't add a substantial overhead in terms of performance.

\textit{stat-weight} performs consistently well across realistic uniform and long-tailed query distributions, it's sensitive to small differences between the rankers and it is robust to noise.

\textit{stat-pruning} performs well in some realistic scenarios, but it felt generally too aggressive and too coupled with the hyper-parameter $\alpha$ that can be tricky to tune.

We validated the intuitions of our analysis and our proposed methods using experiments based on a simulation framework developed from scratch.

Applying \textit{stat-weight} to other interleaved methods in real-world scenarios is an interesting direction for future works.
Also calculating the query credit with different statistical approaches and normalization could be explored.
Finally, it would be interesting to run experiments with bigger numbers and many seeds to see how the different evaluation methods perform.

\appendix

\section{Experiment framework code}

\subsection{Github Repository}

\href{https://github.com/SeaseLtd/statistical-interleaving}{https://github.com/SeaseLtd/statistical-interleaving}

\subsection{Datasets}

\href{http://research.microsoft.com/en-us/projects/mslr/default.aspx.}{$MSLR$}
\href{https://github.com/SeaseLtd/statistical-interleaving/tree/main/data/long-tail-1}{\textit{long-tail-1}}

\subsection{Runs}

Instructions to replicate the runs are in the README.

The detailed output of each run is in:
\href{https://github.com/SeaseLtd/statistical-interleaving/tree/main/runs_output}{output}

%
%
%
\bibliographystyle{splncs04}
\bibliography{samplepaper}

\end{document}